\documentclass[journal]{IEEEtran}

\usepackage{algorithm,algpseudocode}
\usepackage{amstext,amsmath,amssymb,exscale,mathtools,bm,gensymb,cite}
\usepackage{psfrag} 
\usepackage[none]{hyphenat}
\usepackage{tabularx}
\usepackage[shortlabels]{enumitem}
\usepackage{xcolor}
\usepackage[font=small]{caption}
\usepackage{subcaption,siunitx}
\usepackage{scalerel}
\usepackage{makecell}
\usepackage{graphicx}
\usepackage[nolist]{acronym}   
\usepackage{hyperref}
   
\begin{document}

\renewcommand{\baselinestretch}{0.90}
 
\begin{acronym}[MPC] % Give the longest label here so that the list is nicely aligned
	
	\acro{6G}{sixth generation} 
	
	\acro{AR}{auto-regressive}
	
	\acro{AoA}{angle of arrival}
	
	\acro{AoD}{angle of departure}
	
	\acro{AWGN}{additive white Gaussian noise}
	
	\acro{B5G}{beyond fifth generation}
	
	\acro{BS}{base station}
	
	\acro{BPSK}{binary phase-shift keying}
	
	\acro{IRS}{intelligent reflecting surface}
	
    \acro{mmWave}{millimeter Wave}

	\acro{MIMO}{multiple input multiple output}
	
	\acro{CSI}{channel state information}
	
	\acro{DFT}{discrete Fourier transform}
	
    \acro{HOSVD}{high order single value decomposition}
    
	\acro{LOS}{line of sight}
    
    \acro{MMSE}{minimum mean squared error}

    \acro{NMSE}{normalized mean square error}

    \acro{NLOS}{non-line of sight}

	\acro{SNR}{signal-to-noise ratio}
	
	\acro{SVD}{singular value decomposition}
	
	\acro{SE}{spectral efficiency}
	
	\acro{UE}{user equipment}
	
	\acro{ULA}{uniform linear array}
	
	\acro{URA}{uniform rectangular array}
	
	% Algorithms
	
	\acro{ALS}{alternating least squares}
	
	\acro{BALS}{bilinear alternating least squares}
	
	\acro{KF}{Kronecker factorization}
	
	\acro{KRF}{Khatri-Rao factorization}
	
	\acro{LS}{least squares}    
	
	\acro{PARKRON}{PARAFAC-Khatri-Rao-Kronecker factorization}
	
	\acro{TBT}{Tucker-based tracking}
	
\end{acronym}

%\markboth{Journal of \LaTeX\ Class Files,~Vol.~18, No.~9, September~2020}%
%{How to Use the IEEEtran \LaTeX \ Templates}

\title{Tensor-based modeling/estimation of static channels in IRS-assisted MIMO systems}
%\author{Kenneth B. dos A. Benício, Bruno Sokal, Fazal-E-Asim, Andr\'{e} L. F. de Almeida, \\ Behrooz Makki and Gábor Fodor 
\author{Kenneth B. A. Benício, André L. F. de Almeida, Bruno Sokal, Fazal-E-Asim, Behrooz Makki, and Gabor Fodor

\thanks{This work was supported by Ericsson Research, Sweden,
and Ericsson Innovation Center, Brazil, under UFC.51 Technical Cooperation Contract Ericsson/UFC. This study was financed in part by CAPES/Brazil - Finance Code 001, and CAPES/PRINT Proc. 88887.311965/2018-00. André L. F. de Almeida thanks CNPq for its financial support under grant 312491/2020-4. G. Fodor was partially
supported by the Digital Futures project PERCy..}}

%\date{}
    
\maketitle
 
\begin{abstract}
    This paper proposes a tensor-based parametric modeling and estimation framework in \acs{MIMO} systems assisted by intelligent reflecting surfaces (IRSs). We present two algorithms that exploit the tensor structure of the received pilot signal to estimate the concatenated channel. The first one is an iterative solution based on the alternating least squares algorithm. In contrast, the second method provides closed-form estimates of the involved parameters using the high order single value decomposition. Our numerical results  show that our proposed tensor-based methods provide improved performance compared to competing state-of-the-art channel estimation schemes, thanks to the exploitation of the algebraic tensor structure of the combined channel without additional computational complexity.
\end{abstract}


\section{Introduction}  
    Over the last few years, \ac{IRS} has been considered one of the possible technologies to be deployed \ac{B5G} and \ac{6G} wireless networks due to its potential to improve the coverage \cite{rajatheva2021scoring,zheng2022survey}. An \ac{IRS} is a $2$D panel composed of many passive reflecting elements whose elements are capable of independently changing the phase shifts of impinging electromagnetic waves to maximize the \ac{SNR} at the intended receiver \cite{gong2020toward}. Hence, channel estimation must be performed at the end nodes of the network and the receiver should estimate the involved channels from the received pilots reflected by the \ac{IRS} according to a training protocol. Several works have addressed this problem, as mentioned in \cite{pan2022overview,lin2021channel,fazaleasim2023twodimensional,zheng2022compressed,de2021channel,de2022semi,10137372,benicio2023tensorbased}. 
    
    As pointed out in \cite{pan2022overview}, in \ac{IRS}-assisted networks channel estimation methods can be divided into structured and unstructured techniques exploiting the parametric (geometric) modeling of the cascaded channel and methods exploiting the combined channel structure, respectively. Also, in \cite{lin2021channel} a twin-\ac{IRS} structure consisting of two \ac{IRS}s is proposed as a way to obtain the spatial signatures of the involved channels. Also, authors in \cite{fazaleasim2023twodimensional} propose a low-complexity channel parameter estimation that exploits the decoupling of the pilot design along the horizontal and vertical domains.
    
    The authors in \cite{zheng2022compressed} propose channel parameter estimation using a low-rank PARAFAC tensor in the context of \ac{mmWave} systems. The authors in \cite{de2021channel} use a tensor approach to perform supervised channel estimation, in which the decoupling of the \acs{BS}-\ac{IRS} and \ac{IRS}-\acs{UE} channels is achieved. Then, \cite{de2022semi} proposes a tensor-based receiver formulated as a semi-blind problem that jointly estimates the involved channels and transmitted data. The work in \cite{10137372} proposes a set of two tensor-based algorithms to do channel parameter estimation under unknown \ac{IRS} hardware impairments. In our previous work \cite{benicio2023tensorbased}, we propose a two-stage tensor-based framework for parametric channel parameter estimation and data detection of time-varying channels based on a $4$th order PARAFAC model. 
    
    In this paper, we propose a new signal modeling that exploits the geometric channel structure to estimate the spatial signatures of the \ac{IRS}-assisted \ac{MIMO} communication systems and formulate a $3$rd order Tucker tensor model, and derive a set of two tensor algorithms that solve the channel parameter estimation problem by either \ac{ALS} or \ac{HOSVD}. Furthermore, we also study the computational complexity of the proposed schemes and selected benchmark solutions. Our simulation results show that the proposed techniques outperform the classic \ac{LS} and the state-of-the-art \ac{KRF} \cite{de2021channel} algorithms without increasing the computational complexity.

    \textit{Notation}: Scalars, vectors, matrices, and tensors are represented as $a, \boldsymbol{a}, \boldsymbol{A}$, and $\mathcal{A}$. Also, $\boldsymbol{A}^{*}$, $\boldsymbol{A}^{\text{T}}$, $\boldsymbol{A}^{\text{H}}$, and $\boldsymbol{A}^{\dagger}$ stand for the conjugate, transpose, Hermitian, and pseudo-inverse, of a matrix $\boldsymbol{A}$, respectively. The $j$th column of $\boldsymbol{A} \in \mathbb{C}^{I \times J}$ is denoted by $\boldsymbol{a}_{j} \in \mathbb{C}^{I \times 1}$. The operator vec$(\cdot)$ transforms a matrix into a vector by stacking its  columns, e.g., $\text{vec}(\boldsymbol{A}) = \boldsymbol{a} \in \mathbb{C}^{IJ \times 1}$, while the unvec$(\cdot)_{{I \times J}}$ operator undo the operation. The operator D$(\cdot)$ converts a vector into a diagonal matrix,  $\text{D}_j(\boldsymbol{B})$ forms a diagonal matrix $R \times R$ out of the $j$th row of $\boldsymbol{B} \in \mathbb{C}^{J \times R}$. Also, $\boldsymbol{I}_{N}$ denotes an identity matrix of size $N \times N$. The symbols $\otimes$ and $\diamond$ indicate the Kronecker and Khatri-Rao products.

\section{System Model}
    \begin{figure}[!t]
        \centering
        \includegraphics[scale = 0.20]{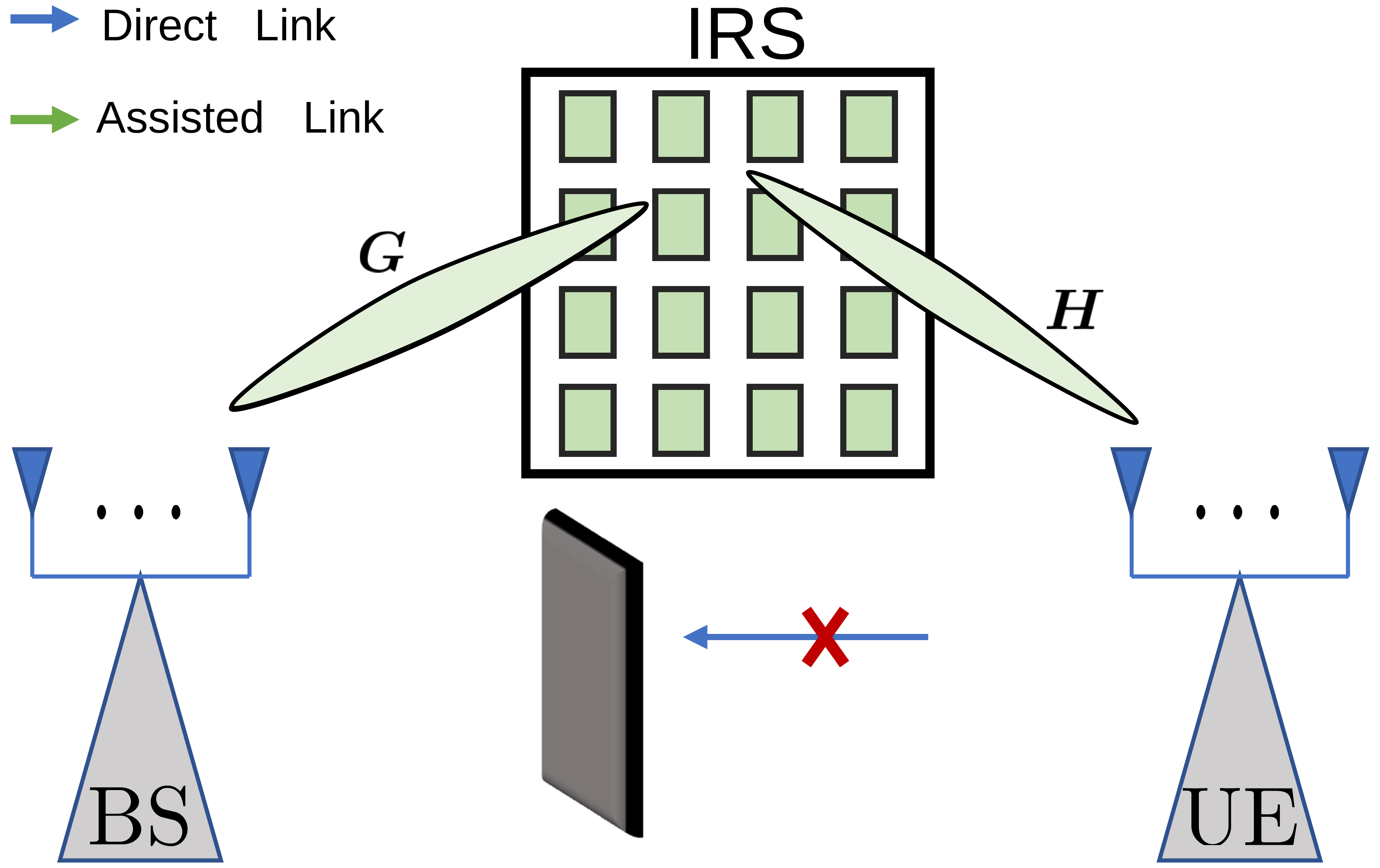}
        \caption{Proposed \acs{IRS}-assisted \acs{MIMO} system scenario.}
        \label{fig:00}
    \end{figure}
    We  consider an uplink \ac{IRS}-assisted \ac{MIMO} scenario with a \ac{BS} equipped with $M$ receiver antennas, which receives a signal from a \ac{UE} equipped with $Q$ transmit antennas \textit{via} a passive \ac{IRS} with $N$ reflecting elements as shown in Fig. \ref{fig:00}. The transmission has length of $T$ time-slots and the received pilot signal at the $t$-th time-slot is given by
    \begin{align}
        \boldsymbol{y}_{t} &\hspace{-0.05cm}=\hspace{-0.05cm} \boldsymbol{G} \text{D}(\boldsymbol{s}_{t}) \boldsymbol{H} \boldsymbol{z}_{t} + \boldsymbol{v}_{t} \in \mathbb{C}^{M \times 1}, \label{signal_model_1}
    \end{align}
    where $\boldsymbol{z}_{t}$ is the pilot sequence, $\text{D}(\boldsymbol{s}_{t})$ is the \ac{IRS} phase-shift matrix, and $\boldsymbol{v}_{t}$ is the \ac{AWGN} vector with $t \in \{1, \cdots, T\}$. We assume that the \ac{IRS}-\ac{UE} channel, $\boldsymbol{H}$, and the \ac{BS}-\ac{IRS} channel, $\boldsymbol{G}$, remain constant during $T$ time-slots and consider a mmWave scenario adopting a multipath channel model \cite{heath2016overview} for the involved channels. We can express these channel matrices as follows
   \begin{align*}
        \boldsymbol{G} &= \sqrt{\frac{K_{G}}{K_{G} + 1}} \boldsymbol{G}^{(\text{LOS})} + \sqrt{\frac{1}{K_{G} + 1}} \boldsymbol{G}^{\text{(NLOS)}}, \\
        \boldsymbol{H} &= \sqrt{\frac{K_{H}}{K_{H} + 1}} \boldsymbol{H}^{(\text{LOS})} + \sqrt{\frac{1}{K_{H} + 1}} \boldsymbol{H}^{\text{(NLOS)}},
    \end{align*}
    \noindent with 
    \begin{align}
        \boldsymbol{G}^{(\text{LOS})} &= \alpha^{(1)} \boldsymbol{a}_{\text{r}_{\text{x}}}(\mu^{(1)}_{\text{bs}})  \boldsymbol{b}^{\text{(irs)}\text{H}}_{\text{t}_{\text{x}}}(\mu^{(1)}_{\text{irs}_{{\text{D}}}},\psi^{(1)}_{\text{irs}_{\text{D}}}), \\
        \boldsymbol{G}^{(\text{NLOS})} &= \sum^{L_{1}}_{l_{1} = 2} \alpha^{(l_{1})} \boldsymbol{a}_{\text{r}_{\text{x}}}(\mu^{(l_{1})}_{\text{bs}})  \boldsymbol{b}^{\text{(irs)}\text{H}}_{\text{t}_{\text{x}}} (\mu^{(l_{1})}_{\text{irs}_{{\text{D}}}},\psi^{(l_{1})}_{\text{irs}_{\text{D}}}), \\
        \boldsymbol{H}^{(\text{LOS})} &= \beta^{(1)} \boldsymbol{b}^{\text{(irs)}}_{\text{r}_{\text{x}}}(\mu^{(1)}_{\text{irs}_{{\text{A}}}},\psi^{(1)}_{\text{irs}_{\text{A}}}) \boldsymbol{a}_{\text{t}_{\text{x}}}^{\text{H}}(\mu^{(1)}_{\text{ue}}), \\
        \boldsymbol{H}^{(\text{NLOS})} &= \sum^{L_{2}}_{l_{2} = 2} \beta^{(l_{2})} \boldsymbol{b}^{\text{(irs)}}_{\text{r}_{\text{x}}}(\mu^{(l_{2})}_{\text{irs}_{{\text{A}}}},\psi^{(l_{2})}_{\text{irs}_{\text{A}}}) \boldsymbol{a}_{\text{t}_{\text{x}}}^{\text{H}}(\mu^{(l_{2})}_{\text{ue}}),
    \end{align}
    where $\boldsymbol{G}^{(\text{LOS})}$, and $\boldsymbol{H}^{(\text{LOS})}$, are the \ac{LOS} components and $\boldsymbol{G}^{(\text{NLOS})}$, and $\boldsymbol{H}^{(\text{NLOS})}$ are the \ac{NLOS} components. Also, $K_{G}$ and $K_{H}$ are the Rician factors for channels $\boldsymbol{G}$ and $\boldsymbol{H}$. 
    The $l$th one-dimensional steering vector of the \ac{BS} is $\boldsymbol{a}_{\text{r}_{\text{x}}}(\mu^{(l_{1})}_{\text{bs}})$ having spatial frequency defined 
    as $\mu^{(l_{1})}_{\text{bs}} = \pi \text{cos}(\phi^{(l_{1})}_{\text{bs}})$ with $\phi^{(l_{1})}_{\text{bs}}$ being the \ac{AoA}, which can be further written in terms of spatial frequency assuming uniform linear array (ULA) as \cite{9298784}
    \begin{align}
        \boldsymbol{a}_{\text{r}_{\text{x}}}(\mu^{(l_{1})}_{\text{bs}}) = \left[1, \cdots, e^{-j\pi (M - 1) \mu^{(l_{1})}_{\text{bs}}}\right]^{\text{T}} \in \mathbb{C}^{M \times 1}.
    \end{align}
    \indent Similarly, the $p$th one-dimensional steering vector for the \ac{UE} is $\boldsymbol{a}_{\text{t}_{\text{x}}}(\mu^{(l_{2})}_{\text{ue}})$ having spatial frequency, which is defined as $\mu^{(l_{2})}_{\text{ue}} = \pi \text{cos}(\phi^{(l_{2})}_{\text{ue}})$, 
    with $\phi^{(l_{2})}_{\text{ue}}$ being the \ac{AoD}, and can be written in terms of spatial frequency as 
    \begin{align}
        \boldsymbol{a}_{\text{t}_{\text{x}}}(\mu^{(l_{2})}_{\text{ue}}) = \left[1, \cdots, e^{-j\pi (Q - 1) \mu^{(l_{2})}_{\text{ue}}}\right]^{\text{T}} \in \mathbb{C}^{Q \times 1}.
    \end{align}
    \indent At the \ac{IRS}, $\boldsymbol{b}^{(\text{irs})}_{\text{r}_{\text{x}}}(\mu^{(l_{2})}_{\text{irs}_{{\text{A}}}},\psi^{(l_{2})}_{\text{irs}_{\text{A}}})$ is the 2D steering 
    vector having spatial frequencies defined as $\mu^{(l_{2})}_{\text{irs}_{{\text{A}}}} = \pi \text{cos}(\phi^{(l_{2})}_{\text{irs}_{\text{A}}}) \text{sin} (\theta^{(l_{2})}_{\text{irs}_{\text{A}}})$
    and $\psi^{(l_{2})}_{\text{irs}_{\text{A}}} = \pi \text{cos}(\phi^{(l_{2})}_{\text{irs}_{\text{A}}})$, where $\phi^{(l_{2})}_{\text{irs}_{\text{A}}}$ and $\theta^{(l_{2})}_{\text{irs}_{\text{A}}}$ 
    are the azimuth \ac{AoA} and the elevation \ac{AoA}, respectively. This can be further written as the Kronecker product between two steering vectors as \cite{9298784}
    \begin{align}
        \boldsymbol{b}^{\text{(irs)}}_{\text{r}_{\text{x}}}(\mu^{(l_{2})}_{\text{irs}_{{\text{A}}}},\psi^{(l_{2})}_{\text{irs}_{\text{A}}}) = \boldsymbol{b}^{\text{(irs)}}_{\text{r}_{\text{x}}}(\mu^{(l_{2})}_{\text{irs}_{{\text{A}}}}) \otimes \boldsymbol{b}^{\text{(irs)}}_{\text{r}_{\text{x}}}(\psi^{(l_{2})}_{\text{irs}_{\text{A}}}) \in \mathbb{C}^{N \times 1}.
    \end{align}
    \indent The \ac{IRS} transmission steering vector, $\boldsymbol{b}^{\text{(irs)\text{H}}}_{\text{t}_{\text{x}}}(\mu^{(l_{1})}_{\text{irs}_{{\text{D}}}},\psi^{(l_{1})}_{\text{irs}_{\text{D}}})$, is defined similarly. The \ac{IRS} phase-shift vector is defined as $\boldsymbol{s}_{t} = \left[e^{j \theta_{t}}, \cdots, e^{j \theta_{N,t}}\right]^{\text{T}}  \in \mathbb{C}^{N \times 1}$, where $\theta_{n,t}$ is the phase-shift of the $n$th \ac{IRS} element at the $t$th time slot. Also, $\boldsymbol{\alpha} = \left[\alpha^{(1)}, \cdots, \alpha^{(L_{1})} \right]^{\text{T}} \in \mathbb{C}^{L_{1} \times 1}$ and $\boldsymbol{\beta} = \left[\beta^{(1)}, \cdots, \beta^{(L_{2})} \right]^{\text{T}} \in \mathbb{C}^{L_{2} \times 1}$ represent the path loss and fading components of the \ac{BS}-\ac{IRS} and \ac{IRS}-\ac{UE} channels, respectively. Both channels are written in matrix notation as
    \begin{align}
        \boldsymbol{G} &= \boldsymbol{A}_{\text{r}_{\text{x}}} \boldsymbol{D} (\boldsymbol{\alpha}) \boldsymbol{B}^{\text{H}}_{\text{t}_{\text{x}}} \in \mathbb{C}^{M \times N}, \label{eq:G} \\
        \boldsymbol{H} &= \boldsymbol{B}_{\text{r}_{\text{x}}} \boldsymbol{D} (\boldsymbol{\beta}) \boldsymbol{A}_{\text{t}_{\text{x}}}^{\text{H}} \in \mathbb{C}^{N \times Q}, \label{eq:H}
    \end{align}
    \noindent where $\boldsymbol{A}_{\text{r}_{\text{x}}}$ and $\boldsymbol{B}_{\text{r}_{\text{x}}}$ are the steering matrices defined as 
    \begin{align*}
        \boldsymbol{A}_{\text{r}_{\text{x}}} &= \left[\boldsymbol{a}_{\text{r}_{\text{x}}}(\mu^{(1)}_{\text{bs}}), \cdots, \boldsymbol{a}_{\text{r}_{\text{x}}}(\mu^{(L_{1})}_{\text{bs}}) \right] \in \mathbb{C}^{M \times L_{1}}, \\
        \boldsymbol{B}_{\text{r}_{\text{x}}} &= \left[\boldsymbol{b}^{(\text{irs})}_{\text{r}_{\text{x}}}(\mu^{(1)}_{\text{irs}_{{\text{A}}}},\psi^{(1)}_{\text{irs}_{\text{A}}}), \cdots, \boldsymbol{b}^{\text{(irs)}}_{\text{r}_{\text{x}}}(\mu^{(L_{2})}_{\text{irs}_{{\text{A}}}},\psi^{(L_{2})}_{\text{irs}_{\text{A}}}) \right] \in \mathbb{C}^{N \times L_{2}},
    \end{align*}
    \noindent with $\boldsymbol{B}_{\text{t}_{\text{x}}} $ and $\boldsymbol{A}_{\text{t}_{\text{x}}}$ being defined in similar manner. 
\section{Pilot-based Parameter Estimation}
    \indent In this section, we describe the proposed tensor-based methods for channel parameter estimation, namely Tucker-ALS as in Alg. \ref{alg:01}, and Tucker-HOSVD as in Alg.\ref{alg:02}. The main idea is to exploit the geometric structure of the involved channels by using a tensor approach.
    \subsection{Tensor-Based Parameter Estimation}  
\begin{figure}[!t]
            \centering
            \includegraphics[scale = 0.10]{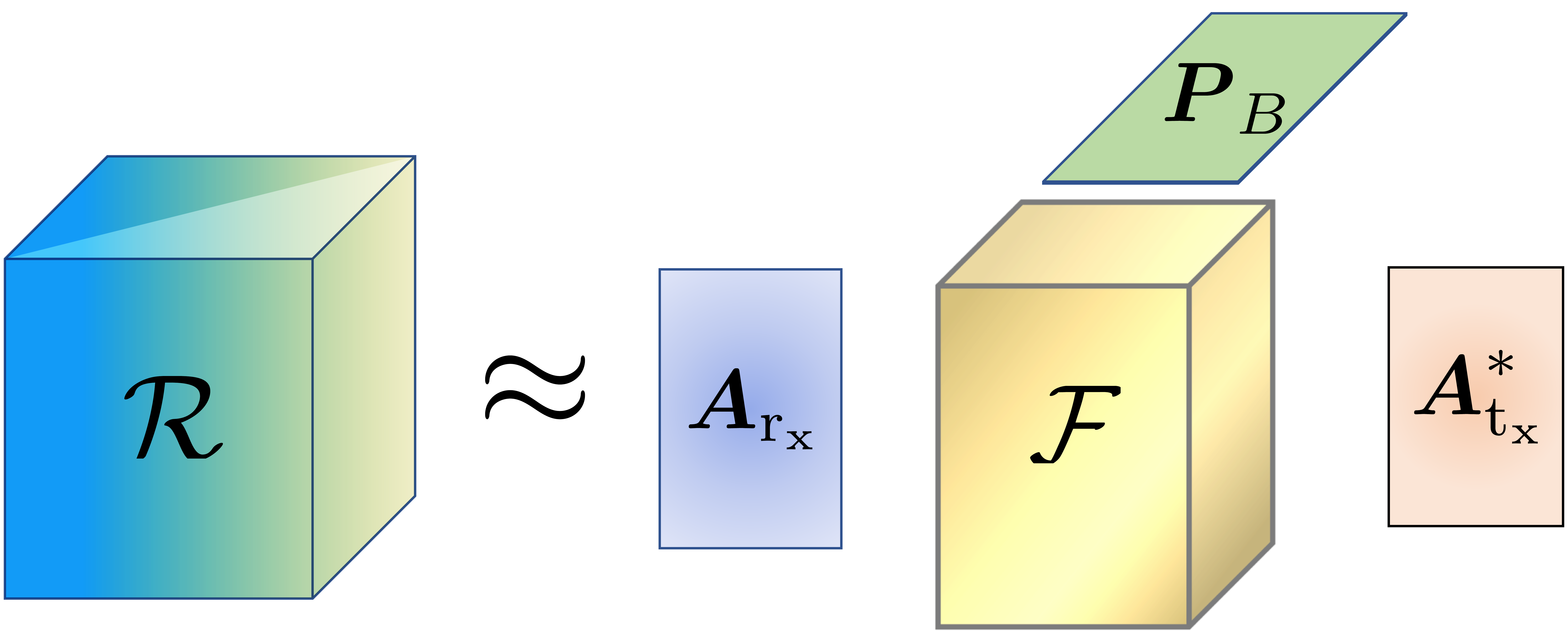}
            \caption{Proposed Tucker-based model.}
            \label{fig:01}
        \end{figure}

        In this section, we formulate a tensor-based approach to estimate the channel parameters. Using $\text{vec}(\boldsymbol{A} \boldsymbol{B} \boldsymbol{C}) = (\boldsymbol{C}^{\text{T}} \otimes \boldsymbol{A}) \text{vec}(\boldsymbol{B})$ and $\text{vec}(\boldsymbol{A} \text{D}\left(\boldsymbol{b}\right) \boldsymbol{C}) = (\boldsymbol{C}^{\text{T}} \diamond \boldsymbol{A})\boldsymbol{b}$ in (\ref{signal_model_1}), yields
        \begin{align}
            \notag \boldsymbol{y}_{t} &= \text{vec }(\boldsymbol{I}_{M} \boldsymbol{G} \text{D}(\boldsymbol{s}_{t}) \boldsymbol{H} \boldsymbol{z}_{t}) + \boldsymbol{v}_{t} \in \mathbb{C}^{M \times 1}, \\
            \notag &= (\boldsymbol{s}^{\text{T}}_{t} \otimes \boldsymbol{z}^{\text{T}}_{t} \otimes \boldsymbol{I}_{M}) \text{vec}(\boldsymbol{H}^{\text{T}} \diamond \boldsymbol{G}) + \boldsymbol{v}_{t}.
        \end{align}
        \begin{align}
            \notag \boldsymbol{y}_{t} &= \text{vec }(\boldsymbol{I}_{M} \boldsymbol{G} \text{D}(\boldsymbol{s}_{t}) \boldsymbol{H} \boldsymbol{z}_{t}) + \boldsymbol{v}_{t} \in \mathbb{C}^{M \times 1}, \\
            \notag &= (\boldsymbol{z}^{\text{T}}_{t} \otimes \boldsymbol{I}_{M}) \text{vec} (\boldsymbol{G} \text{D}(\boldsymbol{s}_{t}) \boldsymbol{H}) + \boldsymbol{v}_{t}, \\
            \notag &= (\boldsymbol{z^{\text{T}}}_{t} \otimes \boldsymbol{I}_{M}) (\boldsymbol{H}^{\text{T}} \diamond\boldsymbol{G}) \boldsymbol{s}_{t} + \boldsymbol{v}_{t},
        \end{align}
        and, applying the first property again, we have
        \begin{align}
            \notag \boldsymbol{y}_{t} &= \text{vec}[(\boldsymbol{z}^{\text{T}}_{t} \otimes \boldsymbol{I}_{M}) (\boldsymbol{H}^{\text{T}} \diamond\boldsymbol{G}) \boldsymbol{s}_{t}] + \boldsymbol{v}_{t}, \\
            \notag &= (\boldsymbol{s}^{\text{T}}_{t} \otimes \boldsymbol{z}^{\text{T}}_{t} \otimes \boldsymbol{I}_{M}) \text{vec}(\boldsymbol{H}^{\text{T}} \diamond \boldsymbol{G}) + \boldsymbol{v}_{t}.
        \end{align}
        \indent Collecting the signals during the $T$ symbol periods yields
        \begin{align}
            \notag \boldsymbol{y} &= \left[ \boldsymbol{y}_{1}^{\text{T}}, \cdots, \boldsymbol{y}_{T}^{\text{T}} \right]^{\text{T}}, \\
            \notag &= [(\boldsymbol{S} \diamond \boldsymbol{Z})^{\text{T}} \otimes \boldsymbol{I}_{M}] \text{vec}(\boldsymbol{H}^{\text{T}} \diamond \boldsymbol{G}) + \boldsymbol{v}, \\
            &= \boldsymbol{\Omega} \hspace{+0.15cm} \boldsymbol{u} + \boldsymbol{v} \in \mathbb{C}^{M T \times 1}, \label{eq:collect}
        \end{align}
        where $\boldsymbol{S} = \left[ \boldsymbol{s}_{1}, \cdots, \boldsymbol{s}_{T} \right] \in \mathbb{C}^{N \times T}$, $\boldsymbol{Z}= \left[ \boldsymbol{z}_{1}, \cdots, \boldsymbol{z}_{T} \right] \in \mathbb{C}^{Q \times T}$ are matrices collecting the \ac{IRS} phase-shifts and pilots, $\boldsymbol{\Omega} = (\boldsymbol{S} \diamond \boldsymbol{Z})^{\text{T}} \otimes \boldsymbol{I}_{M} \in \mathbb{C}^{M T \times M Q N}$,   
        $\boldsymbol{u} = \textrm{vec}(\boldsymbol{H}^{\text{T}} \diamond \boldsymbol{G}) \in \mathbb{C}^{M Q N \times 1}$, and $\boldsymbol{v} = \left[\boldsymbol{v}^{\text{T}}_{1}, \cdots, \boldsymbol{v}^{\text{T}}_{T} \right]^{\text{T}} \in \mathbb{C}^{M T \times 1}$ is the \ac{AWGN} noise term.        
        From (\ref{eq:collect}), we obtain the following \ac{LS} problem
        \begin{align}
            \boldsymbol{\hat{u}} = \underset{\boldsymbol{u}}{\text{arg min}} \left|\left| \boldsymbol{y} - \boldsymbol{\Omega} \boldsymbol{u} \right|\right|^2_{2}, \label{lss}
        \end{align}
        where the solution requires $T \geq Q N$ and is given by
        \begin{align}
            \boldsymbol{\hat{u}} &= \boldsymbol{\Omega}^{\dagger} \boldsymbol{y} \in \mathbb{C}^{MQN \times 1}. \label{eq:03}
        \end{align}
        Let us define $\boldsymbol{R}=\textrm{unvec}_{MQ \times N}(\boldsymbol{\hat{u}}) \approx \boldsymbol{H}^{\text{T}} \diamond \boldsymbol{G} \in \mathbb{C}^{MQ \times N}$, where the approximation is exact in a noiseless scenario. Using (\ref{eq:G}) and (\ref{eq:H}), while applying property $(\boldsymbol{AC}) \diamond (\boldsymbol{BD}) = (\boldsymbol{A} \otimes \boldsymbol{B})(\boldsymbol{C} \diamond \boldsymbol{D})$, we have
        \begin{align}
            \notag \boldsymbol{R} &\approx [\boldsymbol{A}_{\text{t}_{\text{x}}}^{*} \text{D} (\boldsymbol{\beta} ) \boldsymbol{B}^{\text{T}}_{\text{r}_{\text{x}}}] \diamond [\boldsymbol{A}_{\text{r}_{\text{x}}}\text{D} (\boldsymbol{\alpha}) \boldsymbol{B}^{\text{H}}_{\text{t}_{\text{x}}}], \\
            %\notag &\approx (\boldsymbol{A}_{\text{t}_{\text{x}}}^{*} \otimes \boldsymbol{A}_{\text{r}_{\text{x}}}) [(\text{D} (\boldsymbol{\beta}) \boldsymbol{B}^{\text{T}}_{\text{r}_{\text{x}}}) \diamond (\text{D} (\boldsymbol{\alpha}) \boldsymbol{B}^{\text{H}}_{\text{t}_{\text{x}}})], \\
            &\approx \hspace{-0.075cm} (\hspace{-0.030cm} \boldsymbol{A}_{\text{t}_{\text{x}}}^{*} \hspace{-0.075cm} \otimes \hspace{-0.075cm} \boldsymbol{A}_{\text{r}_{\text{x}}}\hspace{-0.030cm} ) \hspace{-0.05cm} [\text{D} (\hspace{-0.025cm}\boldsymbol{\beta}\hspace{-0.025cm}) 
            \hspace{-0.075cm} \otimes \hspace{-0.075cm} \text{D} (\hspace{-0.025cm}\boldsymbol{\alpha}\hspace{-0.025cm})] 
            (\boldsymbol{B}^{\text{T}}_{\text{r}_{\text{x}}} \hspace{-0.075cm} \diamond \hspace{-0.075cm} \boldsymbol{B}^{\text{H}}_{\text{t}_{\text{x}}}). \label{eq:04}
        \end{align}
        \indent Defining $\boldsymbol{f} = \boldsymbol{\beta} \otimes \boldsymbol{\alpha} \in \mathbb{C}^{L_{1} L_{2} \times 1}$, (\ref{eq:04}) can be expressed as
        \begin{align}
            \boldsymbol{R} &\approx (\boldsymbol{A}_{\text{t}_{\text{x}}}^{*} \otimes \boldsymbol{A}_{\text{r}_{\text{x}}}) \text{D}(\boldsymbol{f}) \boldsymbol{P}^{\text{T}}_{B} \in \mathbb{C}^{M Q \times N}, \label{eq:05}
        \end{align}
        \noindent where $\boldsymbol{P}_{B} = (\boldsymbol{B}^{\text{T}}_{\text{r}_{\text{x}}} \diamond \boldsymbol{B}^{\text{H}}_{\text{t}_{\text{x}}}) \in \mathbb{C}^{N \times L_{1} L_{2}}$ is the \ac{IRS} geometry information. Note that $\text{D}(\boldsymbol{f}) \in \mathbb{C}^{L_{1} L_{2} \times L_{1} L_{2}}$ can be viewed as the $3$-mode unfolding of the tensor $\mathcal{F} \in \mathbb{C}^{L_{1} \times L_{2} \times L_{1} L_{2}}$, i.e., $\text{D} (\boldsymbol{f}) = \left[\mathcal{F}\right]_{(3)}$. The tensor $\mathcal{F}$ is given by \cite{sokal2021tensor}
        \begin{align}
            \mathcal{F} &= \left(\mathcal{I}_{3,L_{2}} \otimes^{2,3}_{2,3} \mathcal{I}_{3,L_{1}}\right) \times_{3} \boldsymbol{f}^{\text{T}} \in \mathbb{C}^{L_{1} \times L_{2} \times L_{1} L_{2}}, \label{eq:core}
        \end{align}
        where $\mathcal{I}_{3,L_{2}}$ and $\mathcal{I}_{3,L_{1}}$ are identity tensors and $\otimes^{2,3}_{2,3}$ is the selective Kronecker product (SKP) \cite{sokal2021tensor}, from which we arranged (\ref{eq:05}) following a third-way Tucker tensor structure $\mathcal{R} \in \mathbb{C}^{M \times Q \times N}$, as shown in Fig. \ref{fig:01}, in terms of $n$-mode product as
        \begin{align}
            \mathcal{R} &\approx \mathcal{F} \times_{1} \boldsymbol{A}_{\text{r}_{\text{x}}} \times_{2} \boldsymbol{A}^{*}_{\text{t}_{\text{x}}} \times_{3} \boldsymbol{P}_{B}, \label{tensor}
        \end{align}
        with the matrix unfoldings of $\mathcal{R}$ given by
        \begin{align}
            \notag \left[\mathcal{R}\right]_{(1)} &\approx \boldsymbol{A}_{\text{r}_{\text{x}}} \left[\mathcal{F}\right]_{(1)} (\boldsymbol{P}_{B} \otimes \boldsymbol{A}^{*}_{\text{t}_{\text{x}}})^{\text{T}} \in \mathbb{C}^{M \times Q N}, \\
            \notag \left[\mathcal{R}\right]_{(2)} &\approx \boldsymbol{A}^{*}_{\text{t}_{\text{x}}} \left[\mathcal{F}\right]_{(2)} (\boldsymbol{P}_{B} \otimes \boldsymbol{A}_{\text{r}_{\text{x}}})^{\text{T}} \in \mathbb{C}^{Q \times M N}, \\
            \notag \left[\mathcal{R}\right]_{(3)} &\approx \boldsymbol{P}_{B} \left[\mathcal{F}\right]_{(3)} (\boldsymbol{A}^{*}_{\text{t}_{\text{x}}} \otimes \boldsymbol{A}_{\text{r}_{\text{x}}})^{\text{T}} \in \mathbb{C}^{N \times M Q}.
        \end{align}
        \noindent Consequently, the estimation of $\boldsymbol{A}_{\text{r}_{\text{x}}}$, $\boldsymbol{A}_{\text{t}_{\text{x}}}$, $\boldsymbol{P}_{B}$ and $\mathcal{F}$ consists of solving the following problem
        \begin{equation}
            \left\{\hat{\boldsymbol{A}}_{\text{r}_{\text{x}}}, \hat{\boldsymbol{A}}_{\text{t}_{\text{x}}}, \hat{\boldsymbol{P}_{B}}, \hat{\mathcal{F}}\right\} \hspace{-0.05cm}=\hspace{-0.05cm} \underset{\boldsymbol{A}_{\text{r}_{\text{x}}}, \boldsymbol{A}_{\text{t}_{\text{x}}}, \boldsymbol{P}_{B}, \mathcal{F}} {\text{arg min}} \left|\left|\hspace{-0.05cm} \begin{split} \mathcal{R} -  \mathcal{F} \times_{1} \boldsymbol{A}_{\text{r}_{\text{x}}} \\ \times_{2} \boldsymbol{A}^{*}_{\text{t}_{\text{x}}} \times_{3} \boldsymbol{P}_{B} \end{split} \right|\right|^{2}_{\text{F}}, \label{ten_est}
        \end{equation}           
        which can be performed by means of the well-known \ac{ALS} algorithm \cite{comon2009tensor} or the \ac{HOSVD} algorithm \cite{kolda2009tensor}.
    \subsection{Tucker \acs{ALS}}
        From (\ref{tensor}) we can derive an iterative solution based on the well-known \ac{ALS} algorithm \cite{comon2009tensor}. Here, the algorithm consists of an estimate $\boldsymbol{A}_{\text{r}_{\text{x}}}$, $\boldsymbol{A}_{\text{t}_{\text{x}}}$, $\boldsymbol{P}_{B}$, and $\mathcal{F}$ in an alternating way by iteratively solving the following cost functions
        \begin{align}
            \hat{\boldsymbol{A}}_{\text{r}_{\text{x}}} &= \underset{\boldsymbol{A}_{\text{r}_{\text{x}}}} {\text{arg min}} \left|\left|  \left[\mathcal{R}\right]_{(1)} - \boldsymbol{A}_{\text{r}_{\text{x}}} \left[\mathcal{F}\right]_{(1)} (\boldsymbol{P}_{B} \otimes \boldsymbol{A}^{*}_{\text{t}_{\text{x}}})^{\text{T}}  \right|\right|^{2}_{\text{F}}, \label{eq:ls_1} \\
            \hat{\boldsymbol{A}}_{\text{t}_{\text{x}}} &= \underset{\boldsymbol{A}_{\text{t}_{\text{x}}}} {\text{arg min}} \left|\left|  \left[\mathcal{R}\right]_{(2)} - \boldsymbol{A}^{*}_{\text{t}_{\text{x}}} \left[\mathcal{F}\right]_{(2)} (\boldsymbol{P}_{B} \otimes \boldsymbol{A}_{\text{r}_{\text{x}}})^{\text{T}}  \right|\right|^{2}_{\text{F}}, \label{eq:ls_2} \\
            \hat{\boldsymbol{P}}_{B} &= \underset{\boldsymbol{P}_{B}} {\text{arg min}} \left|\left|  \left[\mathcal{R}\right]_{(3)} - \boldsymbol{P}_{B} \left[\mathcal{F}\right]_{(3)} (\boldsymbol{A}^{*}_{\text{t}_{\text{x}}} \otimes \boldsymbol{A}_{\text{r}_{\text{x}}})^{\text{T}}  \right|\right|^{2}_{\text{F}}, \label{eq:ls_3} \\
            \hat{\boldsymbol{f}} \hspace{-0.05cm}&=\hspace{-0.05cm} \underset{\boldsymbol{f}} {\text{arg min}} \left|\left| \text{vec}(\left[\mathcal{R}\right]_{(3)}) \hspace{-0.05cm}-\hspace{-0.05cm} [(\boldsymbol{A}^{*}_{\text{t}_{\text{x}}} \hspace{-0.05cm}\otimes\hspace{-0.05cm} \boldsymbol{A}_{\text{r}_{\text{x}}}) \hspace{-0.05cm}\diamond\hspace{-0.05cm} \boldsymbol{P}_{B}] \boldsymbol{f}  \right|\right|^{2}_{2}, \label{eq:ls_4}
        \end{align} 
        where $\boldsymbol{f}$ is related to the diagonal elements of $\left[\mathcal{F}\right]_{(3)}$ and its expression is derived by applying $\text{vec}(\boldsymbol{A} \text{D}\left(\boldsymbol{b}\right) \boldsymbol{C}) = (\boldsymbol{C}^{\text{T}} \diamond \boldsymbol{A})\boldsymbol{b}$ to $\left[\mathcal{R}\right]_{(3)}$. The solutions for (\ref{eq:ls_1})-(\ref{eq:ls_4}) are respectively given by
        \begin{align}
            \hat{\boldsymbol{A}}_{\text{r}_{\text{x}}} &=  \left[\mathcal{R}\right]_{(1)} \left[\left[\mathcal{F}\right]_{(1)} (\boldsymbol{P}_{B} \otimes \boldsymbol{A}^{*}_{\text{t}_{\text{x}}})^{\text{T}}\right]^{\dagger}, \label{eq:mode1} \\
            \hat{\boldsymbol{A}}_{\text{t}_{\text{x}}} &=  \left[\mathcal{R}\right]_{(2)} \left[\left[\mathcal{F}\right]_{(2)} (\boldsymbol{P}_{B} \otimes \boldsymbol{A}_{\text{r}_{\text{x}}})^{\text{T}}\right]^{\dagger}, \label{eq:mode2} \\
            \hat{\boldsymbol{P}}_{B} &=  \left[\mathcal{R}\right]_{(3)} \left[\left[\mathcal{F}\right]_{(3)} (\boldsymbol{A}^{*}_{\text{t}_{\text{x}}} \otimes \boldsymbol{A}_{\text{r}_{\text{x}}})^{\text{T}}\right]^{\dagger}, \label{eq:mode3} \\
            \hat{\boldsymbol{f}} &= \left[(\boldsymbol{A}^{*}_{\text{t}_{\text{x}}} \hspace{-0.05cm}\otimes\hspace{-0.05cm} \boldsymbol{A}_{\text{r}_{\text{x}}}) \hspace{-0.05cm}\diamond\hspace{-0.05cm} \boldsymbol{P}_{B}\right]^{\dagger} \text{vec}(\left[\mathcal{R}\right]_{(3)}), \label{eq:corevec}
        \end{align}
        with each solution requiring that $Q N \geq L_{1}$, $M N \geq L_{2}$, $M Q \geq L_{1} L_{2}$, and $N M Q \geq L_{1} L_{2}$. These four conditions are related to the convergence of the \ac{LS} estimates $\boldsymbol{A}_{\text{r}_{\text{x}}}$, $\boldsymbol{A}^{*}_{\text{t}_{\text{x}}}$, $\boldsymbol{P}_{B}$, and $\boldsymbol{f}$, respectively. The proposed \ac{ALS} algorithm consists of four iterative and alternating update steps that follow the \ac{LS} solutions (\ref{eq:mode1})-(\ref{eq:corevec}). At each update, the reconstruction error is minimized according to one given factor matrix by fixing the other matrices to their estimation obtained at the previous update. This procedure is repeated until the convergence is acknowledged, which happens when the reconstruction error, given by $e(i) = ||\mathcal{R} - \hat{\mathcal{R}}(i)||^{2}_{\text{F}}$, achieves $||e(i) - e(i-1)|| \leq \epsilon$ and $\epsilon$ is the threshold parameter with $\hat{\mathcal{R}}(i)$ being the reconstruct tensor model at the $i$th iteration. In this work, we initialize the factor matrices randomly and the convergence threshold is set to $\epsilon = 10^{-5}$. It is worth noting that our solution using the \ac{ALS} is not unique although, after the convergence, the intrinsic scaling and permutation ambiguities disappear on the estimated tensor model $\hat{\mathcal{R}} = \hat{\mathcal{F}} \times_{1} \hat{\boldsymbol{A}}_{\text{r}_{\text{x}}} \times_{2} \hat{\boldsymbol{A}}^{*}_{\text{t}_{\text{x}}} \times_{3} \hat{\boldsymbol{P}}_{B}$.
        \begin{algorithm}[!t]
            \small
            \caption{\acs{ALS} \label{alg:01}}
            \begin{algorithmic}[1]
                \Require{Tensor $\mathcal{R}$}
                \While{$||e(i) - e(i-1)|| \geq \delta$}
                    \State{Find a \ac{LS} estimate of $\boldsymbol{A}_{\text{r}_{\text{x}}}$ with (\ref{eq:mode1}).}
                    \State{Find a \ac{LS} estimate of $\boldsymbol{A}_{\text{t}_{\text{x}}}$ with (\ref{eq:mode2}).}
                    \State{Find a \ac{LS} estimate of $\boldsymbol{P}_{B}$ with (\ref{eq:mode3}).}
                    \State{Find an estimate of $\boldsymbol{f}$ with (\ref{eq:corevec}).}
                    \State{Build $\mathcal{F}$ with (\ref{eq:core}).}
                    \State{Repeat \textit{until} convergence.}
                \EndWhile
                \State \textbf{return} $\hat{\mathcal{R}} = \hat{\mathcal{F}} \times_{1} \hat{\boldsymbol{A}}_{\text{r}_{\text{x}}} \times_{2} \hat{\boldsymbol{A}}^{*}_{\text{t}_{\text{x}}} \times_{3} \hat{\boldsymbol{P}}_{B}$
            \end{algorithmic}
        \end{algorithm}
        
    \subsection{Tucker \acs{HOSVD}}
        Considering the Tucker model in (\ref{tensor}), we can also estimate its factors by finding a multi-linear rank ($L_{1}, L_{2}, L_{1} L_{2}$) approximation to  $\mathcal{R}$. This can be done by means of the state-of-art truncated \ac{HOSVD} algorithm. This algorithm consists of computing multiple singular value decompositions (SVDs), one for each unfolding of  $\mathcal{R}$, as follows
        \begin{align}
            \left[\mathcal{R}\right]_{(1)} &= \boldsymbol{U}^{(1)} \boldsymbol{\Sigma}^{(1)} \boldsymbol{V}^{(1)\text{H}}, \label{eq:svd1} \\
            \left[\mathcal{R}\right]_{(2)} &= \boldsymbol{U}^{(2)} \boldsymbol{\Sigma}^{(2)} \boldsymbol{V}^{(2)\text{H}}, \label{eq:svd2}\\
            \left[\mathcal{R}\right]_{(3)} &= \boldsymbol{U}^{(3)} \boldsymbol{\Sigma}^{(3)} \boldsymbol{V}^{(3)\text{H}}. \label{eq:svd3}
        \end{align}
        The estimates of the steering matrices  $\boldsymbol{A}_{\text{r}_{\text{x}}}$, $\boldsymbol{A}_{\text{t}_{\text{x}}}$, and $\boldsymbol{P}_{B}$ are found from the dominant $L_{1}$, $L_{2}$, and $L_{1} L_{2}$ left singular vectors of $\left[\mathcal{R}\right]_{(1)}$, $\left[\mathcal{R}\right]_{(2)}$, and $\left[\mathcal{R}\right]_{(3)}$, respectively
        \begin{align}
            \hat{\boldsymbol{A}}_{\text{r}_{\text{x}}} = \boldsymbol{U}^{(1)}_{.1:L_{1}},
            \hat{\boldsymbol{A}}_{\text{t}_{\text{x}}} = \boldsymbol{U}^{(2)}_{.1:L_{2}}, 
            \hat{\boldsymbol{P}}_{B} = \boldsymbol{U}^{(3)}_{.1:L_{1} L_{2}}. \label{eq:factormatrices}
        \end{align}
        Finally, an estimate of the core tensor in  (\ref{tensor}) is obtained as
        \begin{align}
            \hat{\boldsymbol{f}} &= \left[(\hat{\boldsymbol{A}}^{*}_{\text{t}_{\text{x}}} \hspace{-0.05cm}\otimes\hspace{-0.05cm} \hat{\boldsymbol{A}}_{\text{r}_{\text{x}}}) \hspace{-0.05cm}\diamond\hspace{-0.05cm} \hat{\boldsymbol{P}}_{B}\right]^{\dagger} \text{vec}(\left[\mathcal{R}\right]_{(3)}), \label{eq:corevechat}
        \end{align}
        where         
        $\boldsymbol{f}$. Similar to the \ac{ALS} solution, the \ac{HOSVD} is not unique once any transformation applied to the core tensor does not change the tensor fit \cite{kolda2009tensor} except in the case where the core tensor is known \cite{favier2016nested}.
        \begin{algorithm}[!t]
            \small
            \caption{Tucker \acs{HOSVD} \label{alg:02}}
            \begin{algorithmic}[1]
                \Require{Tensor $\mathcal{R}$}
                \State{Define the \ac{SVD}s of $\left[\mathcal{R}\right]_{(1)}$, $\left[\mathcal{R}\right]_{(2)}$, and $\left[\mathcal{R}\right]_{(3)}$ with (\ref{eq:svd1})-(\ref{eq:svd3}).}
                \State{Compute the estimation of matrices $\boldsymbol{A}_{\text{r}_{\text{x}}}$, $\boldsymbol{A}_{\text{t}_{\text{x}}}$, and $\boldsymbol{P}_{B}$ with (\ref{eq:factormatrices}).}
                \State{Compute the estimation of $\boldsymbol{f} = \text{vec}(\mathcal{F})$ with (\ref{eq:corevechat})}
                \State{Reconstruct $\hat{\mathcal{F}}$ with (\ref{eq:core})}.
                \State \textbf{return} $\hat{\mathcal{R}} = \hat{\mathcal{F}} \times_{1} \hat{\boldsymbol{A}}_{\text{r}_{\text{x}}} \times_{2} \hat{\boldsymbol{A}}^{*}_{\text{t}_{\text{x}}} \times_{3} \hat{\boldsymbol{P}}_{B}$
            \end{algorithmic}
        \end{algorithm}
    
    \subsection{Computational Complexity}
        In Table \ref{tab:complexity}, we describe the computational complexity for the selected benchmark algorithms, \ac{LS} (\ref{eq:03}) and \ac{KRF} \cite{de2021channel}, and our proposed solutions, \ac{ALS} and \ac{HOSVD}. Consider that the pseudo-inverse of a matrix $\boldsymbol{A} \in \mathbb{C}^{I \times J}$, with $I > J$, and its rank-$R$ \ac{SVD} approximation have complexities $\mathcal{O}(I J^{2})$ and $\mathcal{O}(I J R)$, respectively. Since the design of $\boldsymbol{\Omega}$ at (\ref{eq:03}) is orthogonal, we have $\boldsymbol{\Omega}^{\dagger} = \boldsymbol{\Omega}^{\text{H}}$ which lower the cost of the pseudo-inverse computation. The \ac{KRF} \cite{de2021channel} estimates the combined channel, $\boldsymbol{R} = \boldsymbol{H}^{\text{T}} \diamond \boldsymbol{G}$, by finding estimates of both $\boldsymbol{G}$ and $\boldsymbol{H}$ that solves a set of $N$ rank-one approximations using the \ac{SVD}. Regarding the proposed algorithms, the \ac{ALS}  computes $4$ pseudo-inverses (\ref{eq:mode1})-(\ref{eq:corevec}) along $\text{\acs{ALS}}_{\text{iter}}$ iterations until convergence, while the \ac{HOSVD} involves $3$ \ac{SVD}s (\ref{eq:svd1})-(\ref{eq:svd3}) and a pseudo-inverse (\ref{eq:corevechat}).
        % \begin{table}[!t]
        %     \centering
        %     \caption{Algorithm computational complexity}
        %     \label{tab:complexity}
        %     \resizebox{0.475\textwidth}{!}{
        %     \begin{tabular}{|c|c|}
        %     \hline
        %     \bold{Algorithm}                & \bold{Computational Complexity}   \\ \hline
        %     \acs{LS}  (\ref{eq:03})         & $\mathcal{O}((MQN)^{2})$                             \\ \hline
        %     \acs{KRF} \cite{de2021channel}  & $\mathcal{O}(M Q N)$                         \\ \hline
        %     Tucker \acs{ALS}                & $\mathcal{O}(M Q N [\text{\acs{ALS}}_{\text{iter}} (\frac{L^{2}_{1}}{M} + \frac{L^{2}_{2}}{Q} + \frac{(L_{1} L_{2})^{2}}{N} + (L_{1} L_{2})^{2})])$                                                               \\ \hline
        %     Tucker \acS{HOSVD}              & $\mathcal{O}(MQN(L_{1} + L_{2} + L_{1} L_{2} + (L_{1} L_{2})^{2}))$ \\ \hline
        %     \end{tabular}}
        % \end{table}

\section{Simulation Results}
\begin{figure}[!t]
        \centering
        \begin{minipage}{.485\columnwidth}
            \centering
            \includegraphics[width = 0.95\linewidth, height = 4.9cm]{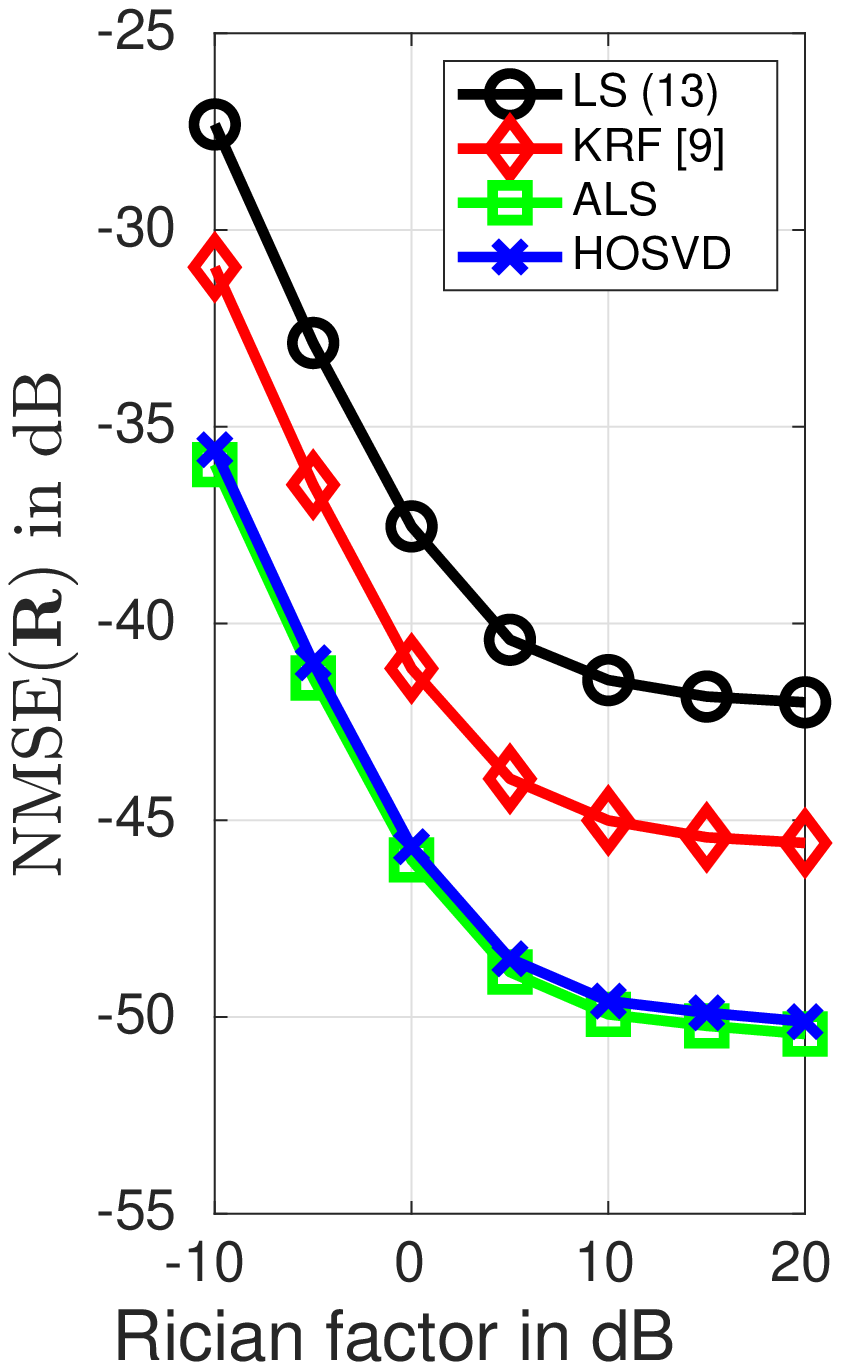}
            \caption{NMSE as a function of the Rician factors $K_{G}$ and $K_{H}$.}
            \label{fig:02}
        \end{minipage}
        \begin{minipage}{.485\columnwidth}
            \centering
            \includegraphics[width = 0.95\linewidth, height = 4.9cm]{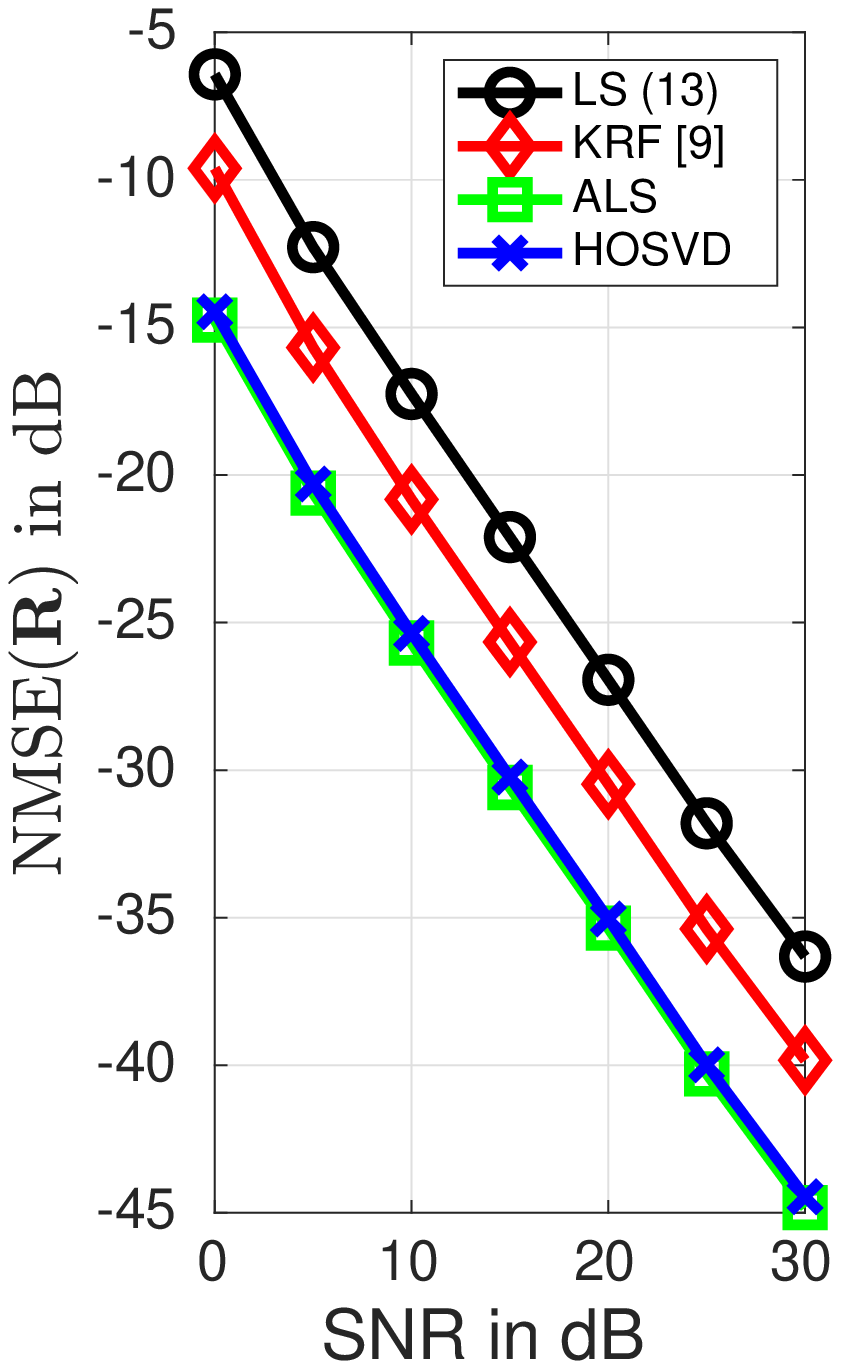}
            \caption{NMSE as a function of the training \ac{SNR}.}
            \label{fig:03}
        \end{minipage}
    \end{figure}

\begin{figure}[!t]
        \centering
        \begin{minipage}{.485\columnwidth}
            \centering
            \includegraphics[width = 0.95\linewidth, height = 4.9cm]{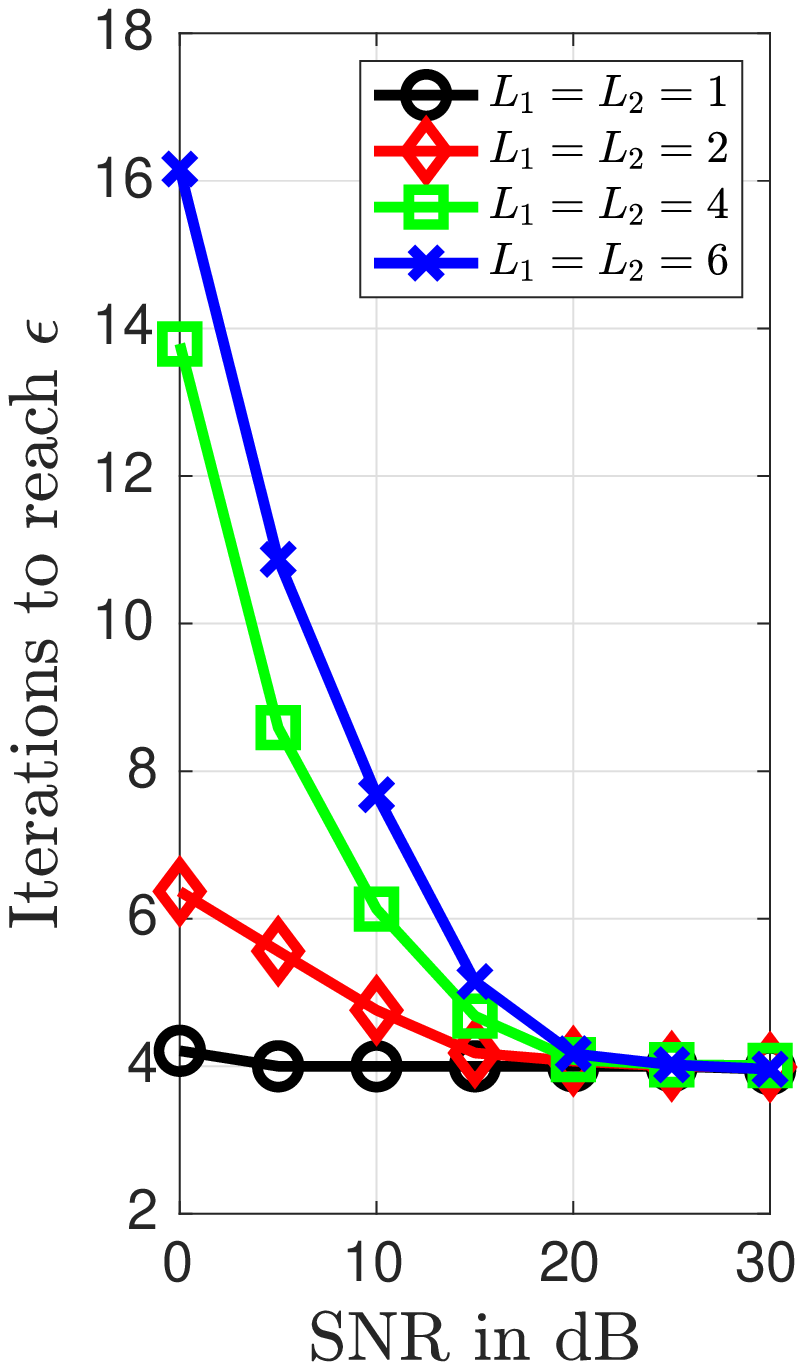}
            \caption{Iterations to converge as function of the \ac{SNR} and for increasing $L_{1}$ and $L_{2}$.}
            \label{fig:04}
        \end{minipage}
        \begin{minipage}{.485\columnwidth}
            \centering
            \includegraphics[width = 0.95\linewidth, height = 4.9cm]{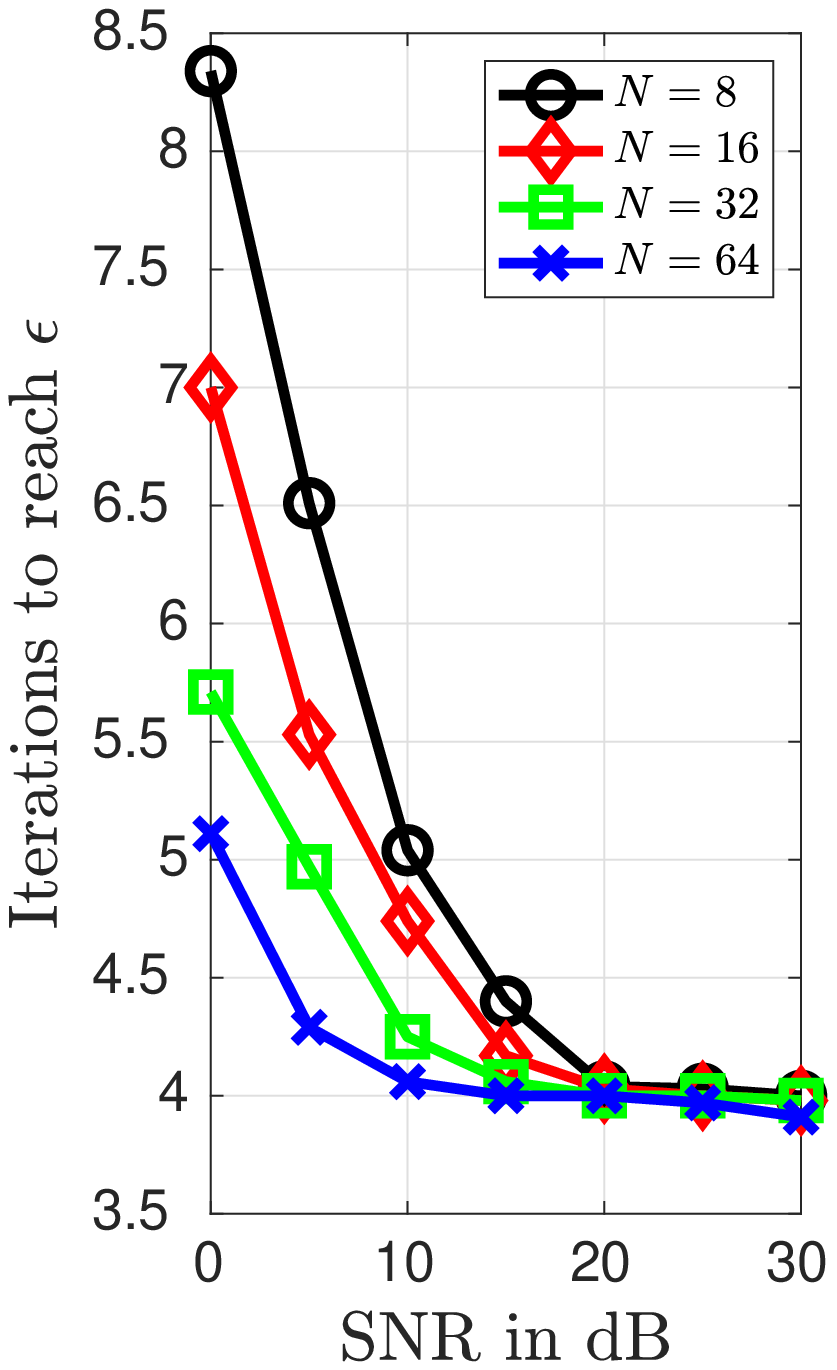}
            \caption{Iterations to converge as function of the \ac{SNR} and for increasing $N$.}
            \label{fig:05}
        \end{minipage}
    \end{figure}

\begin{figure}[!t]
        \centering
        \includegraphics[scale = 0.40]{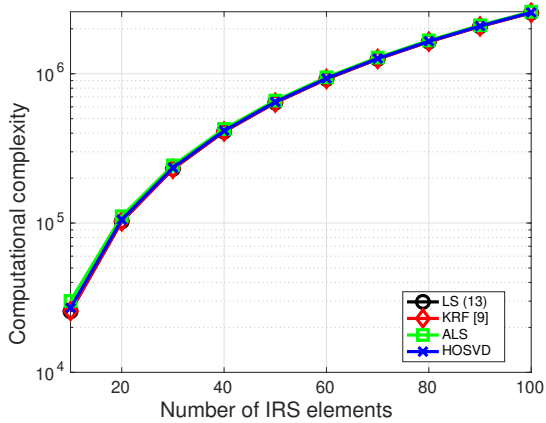}
        \caption{Computational complexity of the proposed solutions and the benchmark algorithms \ac{LS} (\ref{eq:03}) \ac{KRF} \cite{de2021channel}.}
        \label{fig:06}
    \end{figure}

    \indent We evaluate the performance of the proposed tensor-based algorithm by comparing it with the reference parameter estimation method based on the \ac{KRF} \cite{de2021channel}. The pilot matrix $\boldsymbol{Z} \in \mathbb{C}^{Q \times T}$ is designed as a Hadamard matrix, while a \ac{DFT} is adopted for the \ac{IRS} phase-shift matrix $\boldsymbol{S}$. The angular parameters $\phi^{(l_{1})}_{\text{bs}}$ and $\phi^{(l_{2})}_{\text{ue}}$ are randomly generated from a uniform distribution between $[-\pi, \pi]$ while the \ac{IRS} elevation and azimuth angles of arrival and departure are
    randomly generated from a uniform distribution between $[-\pi/2, \pi/2]$. The fading coefficients $\boldsymbol{\alpha}$ and $\boldsymbol{\beta}$ are modeled as independent Gaussian random variables $\mathcal{CN}(0,1)$. The parameter estimation accuracy is evaluated in terms of the \ac{NMSE} given by 
    \begin{align}
        \notag &\text{\ac{NMSE}}(\boldsymbol{R}) = \mathbb{E}\left\{\frac{\left|\left|\boldsymbol{R}^{(m)} - \hat{\boldsymbol{R}}^{(m)}\right|\right|^{2}_{\text{F}}}{\left|\left|\boldsymbol{R}^{(m)}\right|\right|^{2}_{\text{F}}}\right\},
    \end{align}
    with $\boldsymbol{R} = (\boldsymbol{H}^{\text{T}} \diamond \boldsymbol{G})$ being the estimated channel at the $m$th experiment, $M = 10^4$ being the number of Monte Carlo experiments, and $\sigma^{2}_{\boldsymbol{V}}$ is the noise variance. Unless otherwise stated, the training \ac{SNR} is $30$ dB, the Rician factor of the \ac{LOS} channel is $K_{G} = 10$ dB, and the Rician factor of the \ac{NLOS} channel is $K_{H} = -10$ dB. At Figs. \ref{fig:02}, \ref{fig:03}, and \ref{fig:06} we assume $\{M = 4, Q = 4, L_{1} = 1, L_{2} = 4, N = 16, \text{ and } T = 64\}$, at Fig. \ref{fig:04} we assume $\{M = 8, Q = 8, N = 16, \text{ and } T = 128\}$, and finally at Fig. \ref{fig:05} we assume $\{M = 8, Q = 8, L_{1} = 2, L_{2} = 2, \text{ and } T = 8N\}$.

    In Fig. \ref{fig:02}, we show the impact of the Rician factors, $K_{G}$ and $K_{H}$, on the \ac{NMSE} performance associated with the estimation of the combined channel $\boldsymbol{R} = \boldsymbol{H}^{\text{T}} \diamond \boldsymbol{G}$, for the proposed algorithms and the benchmark \ac{KRF} \cite{de2021channel}. The performance improvements with increasing values of the Rician factor come from the fact that as $K_{G}$ and $K_{H}$ increases, both channels are dominated by a \ac{LOS} component, which happens because the channel approaches a unitary rank and the noise rejection coming from the \ac{SVD}. We can also see that the proposed algorithms outperform the classic \ac{LS} (\ref{eq:03}) and the state-of-the-art \ac{KRF} \cite{de2021channel} algorithms by approximately $10$ dB and $5$ dB, respectively. Moreover, the \ac{ALS} performance is better than that of the \ac{HOSVD} in about 1 dB. In Fig. \ref{fig:03}, we evaluate the \ac{NMSE} performance as a function of the training \ac{SNR} and verify the same gains as the evaluation of the \ac{NMSE} in terms of the Rician factors. In all considered methods, we observe that the \ac{NMSE} decreases with \ac{SNR} linearly.

    In Fig. \ref{fig:04}, we evaluate the number of iterations required by the proposed \ac{ALS} algorithm to converge as a function of the \ac{SNR} for a varying number of paths, $L_{1}$ and $L_{2}$. In this figure, we define $\epsilon = 10^{-5}$ as the target convergence criterion, meaning that the algorithm convergence is declared when the error between consecutive iterations is less than $\epsilon$. As expected, in the low \ac{SNR} region the \ac{ALS} algorithm takes more iterations to achieve convergence as the total number of components, i.e., the product $L_{1} L_{2}$, increases. At the high \ac{SNR} region, the required number of iterations is the same. In Fig. \ref{fig:05}, we observe that, as the number of reflecting elements $N$ increases, fewer iterations are needed for convergence. This is linked to \ac{LS} (\ref{eq:03}) since, if $N$ increases, we sense the channel longer because of condition $T \geq QN$.

    In Fig. \ref{fig:06}, we analyze the computational complexity of the benchmark algorithms, \ac{LS} (\ref{eq:03}) and \ac{KRF} \cite{de2021channel}, and the proposed algorithms, \ac{ALS} and \ac{HOSVD}, for fixed parameters $M, Q, L_{1}, L_{2}$ while varying $N$ according to Table \ref{tab:complexity}. To compute the cost of the \ac{KRF} \cite{de2021channel}, the \ac{ALS}, and the \ac{HOSVD}, we take into account the extra cost of the \ac{LS} (\ref{eq:03}) step, which is the most complex operation of the competing and our proposed solutions (see Table \ref{tab:complexity}). We observe that the competing \ac{LS} and \ac{KRF} \cite{de2021channel} algorithms have approximately the same cost as the proposed \ac{ALS} and \ac{HOSVD} solutions.

\section{Conclusions}
     This paper proposes two tensor-based channel parameter estimation algorithms in \ac{IRS}-aided \ac{MIMO} communications. From our simulation results, we observed that proposed \ac{ALS} and \ac{HOSVD} algorithms outperform both the classic \ac{LS} and the state-of-the-art \ac{KRF} algorithms in terms of \ac{NMSE} by approximately $10$ dB and $5$ dB, respectively. The performance gap between the proposed solutions is small with the \ac{ALS} algorithm having the best performance in terms of \ac{NMSE} only. Regarding the computational complexity, the proposed \ac{ALS} and \ac{HOSVD} solutions have approximately the same complexity  as the benchmark ones, namely, the \ac{LS} (\ref{eq:03}) and the \ac{KRF} \cite{de2021channel} methods.

\end{document}